\documentclass[
  aps,
  reprint,
  superscriptaddress,
  floatfix,
  nofootinbib,
  longbibliography
]{revtex4-2}

\usepackage{amsmath,amssymb}
\usepackage{graphicx}
\usepackage{xcolor}
\usepackage{tikz}
\usetikzlibrary{calc,positioning,arrows.meta}
\usepackage{hyperref}

\hypersetup{
  colorlinks=true,
  linkcolor=blue,
  citecolor=blue,
  urlcolor=blue
}

\newcommand{\R}{\mathbb{R}}
\newcommand{\E}{\mathbb{E}}
\newcommand{\Prob}{\mathbb{P}}
\newcommand{\T}{^{\mathsf T}}
\newcommand{\norm}[1]{\lVert #1\rVert}

\begin{document}

\title{Quasi-Orthogonality and Polysemy:\texorpdfstring{\\}{ }
The Limits of Quantum Analogy in Large Language Models}

\author{Karl Svozil}
\email{karl.svozil@tuwien.ac.at}
\affiliation{Institute for Theoretical Physics, TU Wien (Vienna University of Technology),
Wiedner Hauptstra\ss{}e 8--10/136, 1040 Vienna, Austria}

\date{\today}

\begin{abstract}
We analyze two geometric aspects of meaning-sensitive behavior in next-token
models.  First, high-dimensional geometry permits exponentially many nearly
orthogonal feature directions at a fixed tolerance, although their residual
overlaps limit simultaneous readout.  Second, polysemy requires us to
distinguish classical convex mixtures of sense vectors, context-dependent
occurrence vectors, quantum coherence, and an optional shared-projector
incidence construction over sense-labelled bases.  For ordinary transformers,
the quantum comparisons are structural and deliberately limited.
\end{abstract}

\maketitle

\section{From formal prediction to meaning-sensitive behavior}
\label{sec:puzzle}

A large language model (LLM) receives strings and returns strings.  Yet it can
answer questions, continue an argument, translate a passage, and alter its
wording when the surrounding text changes.  Its behavior is therefore
sensitive to distinctions that readers describe as semantic.

The question recalls John Searle's Chinese Room thought experiment
\cite{searl-80a}.  A person who does not know Chinese follows formal rules for
transforming Chinese symbols and thereby produces answers that appear
intelligent to an outside reader.  Searle asks whether successful symbol
manipulation is sufficient for understanding---whether semantics follows from
syntax.  We do not try to settle that philosophical question, and fluent
behavior will not be used as evidence for the emergence of meaning or even
consciousness.  Our question is narrower and operational: what internal
organization allows calculations on tokens to construct and exploit distinctions that
human readers recognize as semantic?

An LLM computation begins by encoding text as vectors.  A tokenizer maps a
string to token identifiers, and an embedding table maps each identifier to a
learned lookup vector.  After positional information is supplied, transformer
layers use the available sequence context---the preceding tokens in a causal
LLM---to construct a context-dependent vector for each token occurrence.

This use of vectors creates a limited formal point of contact with quantum mechanics,
where a pure state is represented by a ray in a complex Hilbert space.
Between measurements, an isolated system evolves unitarily---essentially by one-to-one
reversible linear transformations preserving vector lengths.
A projective measurement stochastically yields one of its possible outcomes
with probabilities given by the Born rule.
In complex Hilbert spaces of dimension at least three, Gleason's theorem~\cite{Gleason,Wright2019}
shows that a normalized, context-independent probability assignment additive over mutually orthogonal projectors
(corresponding to mutually exclusive propositions or outcomes) gives that Born rule.

Against this shared inner-product background, two issues provide a manageable
way to examine the analogy and its limits.

First, a fixed-width representation space of dimension $d$ can contain at
most $d$ mutually orthogonal directions, yet a model may need to distinguish
vastly more properties. By relaxing exact orthogonality to permit small
pairwise overlaps, a high-dimensional space can instead accommodate many more
than $d$ nearly orthogonal directions. We examine the resulting tradeoff:
\emph{quasi-orthogonality} expands representational capacity, but the residual
overlaps accompanying this expansion can accumulate as readout cross-talk,
thereby limiting access to the encoded distinctions.

Second, one word can have several meanings.  The word ``bank'' may refer to a
financial institution, the edge of a river, or a seat.  We ask how a single
token type and its many occurrences can represent such polysemy.  This leads
to comparisons with two different quantum ideas: coherent superposition and
one lexical projector shared by several sense-labelled orthonormal bases.

The order of exposition is as follows:  Section~\ref{sec:vectors} begins
with tokenization and explains why vector representations are useful.
Section~\ref{sec:quasi} develops the abundance--cross-talk tradeoff and a
limited comparison with quantum decoherence.  Section~\ref{sec:polysemy}
distinguishes contextual polysemy from coherent superposition and introduces
the shared-projector construction.  Section~\ref{sec:prior-work} relates these
comparisons to adjacent literatures.  Appendix~\ref{app:quantum-attention},
\emph{A minimal quantum attention model as a representational foil}, makes explicit the extra
structure used in a phase-sensitive quantum construction.  It is not a
reinterpretation of ordinary transformer attention, an efficient quantum
algorithm, or a claim of computational advantage.

\section{From text to vectors}
\label{sec:vectors}

\subsection{Tokenization: symbols before geometry}

The first operation is \emph{tokenization}.  A tokenizer divides a character
string into units drawn from a finite vocabulary and replaces each unit by an
integer identifier.  A token may be a complete word, part of a word,
punctuation, a space-bearing fragment, or a byte-level unit.  Thus a token is
not automatically a word, still less a word's meaning.  Subword schemes
assemble rare and previously unseen words from reusable
fragments~\cite{sennrich2016subword}.

The assigned integer is merely an address.  Identifier 7193 is not
semantically closer to identifier 7194 than to identifier 20.  The identifiers
are therefore used as addresses in a table of vectors on which graded
comparison and learned transformation are possible.

\subsection{A token vector is a starting point, not a complete meaning}

A learned lookup table assigns a real vector to every token identifier.  Two
occurrences of the same identifier begin with the same table row.  The model
must also know order: ``the dog chased the cat'' is not equivalent to ``the
cat chased the dog.''  Position is therefore supplied in some form, either by
adding positional information or by using another positional mechanism.

The resulting initial vector is not a contextually complete meaning.  It has
been trained across many occurrences and can reflect what those occurrences
share, but later layers construct a new hidden vector for each particular
occurrence.  The occurrence of ``bank'' near ``loan'' can consequently acquire
a different hidden state from the same token near ``river,'' although both
began from the same lookup row.  This distinction between a token \emph{type}
and a contextual token \emph{occurrence} will be essential for polysemy.

Successive transformer layers let each token position combine information from
the positions available to it, thereby producing context-dependent vectors.
During training, prediction errors adjust persistent model parameters.  During
ordinary response generation those parameters remain fixed, and each selected
token is appended to the text before the next token is predicted.

\subsection{Why vectors?}

Vectors are useful for several reasons.  Their coordinates can be adjusted
continuously during learning.  Matrices can transform them and inner products
can compare them.  Vector addition permits several contributions to coexist
in one state.  Most importantly here, lengths, directions, projections, and
subspaces provide an economical language for graded features.

Suppose a unit vector $u$ has been fitted as a direction associated with some
property at a chosen layer.  Relative to a declared baseline $h_0$, the scalar
$u\T(h-h_0)$ is the signed coordinate of a hidden state $h$ along that
direction.  It is the coordinate, not the axis itself, that says on which side
of the contrast the state lies and how strongly.  In a simplified gender
example, the one-dimensional subspace $\mathrm{span}\{u\}$ supplies a fitted
direction for a gender-associated contrast.  The subspace is unoriented:
$u$ and $-u$ span the same line.  One may orient $u$ from a feminine-marked
sample mean toward a masculine-marked sample mean.  Together with $h_0$, this
defines the affine axis $h_0+\mathrm{span}\{u\}$.  Positive and negative projections then
describe opposite sides of the same axis; they are not two orthogonal
directions.  The baseline or a probe intercept defines the zero.  Modern
contextual representations may use several related directions rather than one
universal axis, and the score can change from sentence to sentence
\cite{bolukbasi2016gender,zhao2019contextualgender}.  The superscript
$\mathsf T$ means transpose, so $u\T h$ is the ordinary Euclidean inner
product.

This directional interpretation is operational, not metaphysical.  It depends
on the layer, data, centering, inner product, and extraction method.  A probe
that reads a property shows that the information is linearly available; it
does not by itself prove that the model uses that direction causally.
Orthogonality nevertheless has one precise consequence.  If two property
directions are perpendicular, a component along either direction contributes
no cross-talk when the other is read by an inner product.  The two directions
are therefore \emph{geometrically decoupled for this linear readout}.  If one
uses the phrase ``operational semantic independence,'' it means only this
absence of readout cross-talk.  It does not mean that the corresponding concepts are
logically unrelated, statistically independent in the data, or causally
independent inside the model.

The reason to emphasize vectors is therefore not that meanings are known in
advance to be points in a Euclidean space.  It is that a trained vector system
can discover useful graded distinctions, combine them, and compare them by
simple operations.  High dimension adds a striking opportunity: far more
approximately separated directions can coexist than there are exactly
independent coordinates.  That opportunity motivates our first central
issue.

\section{Quasi-orthogonal directions in a finite-dimensional space}
\label{sec:quasi}

\subsection{Exact independence and approximate separation}

In a $d$-dimensional space, no more than $d$ nonzero vectors can be mutually
orthogonal.  This is elementary linear algebra: mutually orthogonal nonzero
vectors are linearly independent, and a linearly independent family cannot
contain more vectors than the dimension.  If semantic distinctions required
exactly perpendicular axes, fixed width would impose a severe limit.

Exact orthogonality is stronger than most readouts require.  Let $u$ and $v$
be unit vectors.  Their inner product is the cosine of the angle between them.
We call them quasi-orthogonal at tolerance $\varepsilon$ when
$|u\T v|\leq\varepsilon$, where the tolerance is chosen before the comparison.
They are not independent coordinates; they are approximately separated
directions.

High-dimensional geometry makes such directions surprisingly abundant.
For a uniformly random unit vector in $d$ real dimensions, the overlap with a
fixed unit direction is typically of size $1/\sqrt d$.  Most of the sphere is
therefore concentrated near the equator perpendicular to any chosen
direction.  L\'evy's concentration principle expresses this typicality.
The Johnson--Lindenstrauss lemma approaches the same phenomenon from another
side: a finite collection of points can be placed in a dimension proportional
to the logarithm of the number of points while approximately preserving all
pairwise distances.  Read in the constructive direction, these results show
that at any fixed nonzero angular tolerance a $d$-dimensional representation
space admits
at least exponentially many quasi-orthogonal directions
\cite{ledoux2001concentration,johnson-1984,Dasgupta-2003}.

The word \emph{exponential} is important, but so are its qualifications.  It
counts directions available at the chosen tolerance, not a second
vector-space dimension.  It is a geometric possibility, not proof that a
trained model uses all those directions.  The allowed overlap must be stated,
and an exponentially large dictionary is not automatically readable when
many of its elements are active at once.  Appendix~\ref{app:vectors}
introduces the notation, and Appendix~\ref{app:geometry} gives the
concentration, union-bound, and Johnson--Lindenstrauss calculations.

\subsection{The price of abundance: readout cross-talk}

The simplest calculation already shows the access problem.  Suppose a hidden
state contains two unit feature directions,
\begin{equation}
 h=a u+b v.
 \label{eq:two-feature-main}
\end{equation}
Reading the first feature by an inner product gives
$u\T h=a+b(u\T v)$.  The desired coefficient is $a$; the second term is
cross-talk from the other feature.  Exact orthogonality removes it, while
quasi-orthogonality bounds its size by $\varepsilon|b|$.

Many small unwanted terms can nevertheless accumulate.  If $k$ comparably
sized, random-like features are active and their signs do not align
systematically, their mean squared errors add as in the Pythagorean theorem.
The resulting root-mean-square (RMS) cross-talk is of order
$\sqrt{k/d}$ relative to one unit-scale target.  This is an average random-code
benchmark, not a guarantee for every learned dictionary.  In the worst case,
aligned cross-terms can add rather than cancel.

The useful operating region is consequently task dependent.  There should be
enough distinguishable directions for the features a task needs, but not so
many simultaneously active, strongly overlapping directions that the chosen
readout fails.  There is no universal numerical boundary determined by $d$
alone.  It also depends on which features coactivate, their strengths, the
decoder, and the tolerated error.  A learned model may use its angular budget
selectively, placing greater separation between features that frequently need
to be read together.

\subsection{Readout cross-talk and a limited quantum comparison}

Small overlaps also occur in quantum theory.  When alternatives of a quantum
system become correlated with nearly orthogonal environmental states, the
off-diagonal terms of the reduced state are multiplied by the corresponding
environmental overlaps.  High dimension can make those factors very small
when the conditional states or their relative dynamics satisfy an explicit
typicality assumption; dimension alone is not enough, since the conditional
states could remain identical.  The author's companion analysis studies this
purely geometric contribution to decoherence
\cite{svozil2026geometricdecoherence}.

For an LLM, quasi-orthogonal feature directions reduce linear-readout
cross-talk.  In environmental decoherence, nearly orthogonal conditional
environmental states reduce off-diagonal terms of the reduced state.  Both
statements contain small inner products, but only the second describes
physical decoherence.  An LLM has no required system--environment split,
unitary branch dynamics, or partial trace.  The common element is algebraic,
not dynamical or physical \cite{RevModPhys.75.715}.

The word \emph{orthogonal} also has different operational meanings here.  For
the LLM readout it means geometric decoupling: absence of linear cross-talk in
the chosen inner product.  Within one quantum measurement
context, orthogonal rays represent
mutually exclusive outcomes.  The common inner-product geometry does not make
these interpretations identical.

This qualified comparison is still useful.  In both settings, exact
orthogonality is an ideal limit; approximate orthogonality can already make
alternatives operationally difficult to confuse; and the number of pairwise
small overlaps is not the same as the number of alternatives that can be
reliably recovered together.

\section{Polysemy: one form, several meanings}
\label{sec:polysemy}

\subsection{Terms}

A \emph{sense} is one conventional meaning of a word form.  For example,
``bank'' may mean a financial institution, the edge of a river, a bench or
long seat, or the lateral tilt of an aircraft.  We use \emph{polysemy} in the
broad computational sense: one word form has more than one sense.

For simplicity, assume that the tokenizer represents ``bank'' by one token.
Its vocabulary entry is the \emph{token type}; one use in a text is a
\emph{token occurrence}.  The surrounding text that helps identify its sense
is the \emph{linguistic context}.  A \emph{semantic feature}, such as animacy,
is one reusable property that may help characterize a sense; it is not itself
a complete word meaning.

\subsection{Static and contextual vectors}

A static embedding such as word2vec assigns the same vector to every
occurrence of a token type \cite{mikolov2013}.  Under the generative model of
Arora and collaborators, this type vector is approximately a
frequency-weighted average of hypothetical sense vectors
\cite{arora-etal-2018-linear}.  Retaining three senses of ``bank'' for the
illustration gives
\begin{equation}
\begin{aligned}
 v_{\rm bank}
 &\approx p_{\rm finance}v_{{\rm bank},{\rm finance}} \\
 &\qquad+p_{\rm river}v_{{\rm bank},{\rm river}}\\
 &\qquad+p_{\rm seat}v_{{\rm bank},{\rm seat}},\\
 &p_{\rm finance},p_{\rm river},p_{\rm seat}\geq 0,\\
 &p_{\rm finance}+p_{\rm river}+p_{\rm seat}=1 .
\end{aligned}
 \label{eq:static-mixture-main}
\end{equation}
Here $v_{{\rm bank},s}$ is the hypothetical vector for sense $s$, and $p_s$
is its relative corpus frequency.  Equation
\eqref{eq:static-mixture-main} is a classical convex combination: it
summarizes a token type over the corpus but does not identify the sense of one
occurrence.  It also does not require the sense vectors to be orthogonal.

A transformer likewise has one initial lookup vector for a token type, but it
then computes a hidden vector for each occurrence from the available text.
Thus the occurrences of ``bank'' in ``central bank'' and ``river bank'' begin
with the same lookup vector but can acquire different hidden vectors.  This
contextualization need not implement the convex sum in
Eq.~\eqref{eq:static-mixture-main}.

\subsection{Why this is not a quantum superposition}

In Eq.~\eqref{eq:static-mixture-main}, the coefficients are nonnegative real
frequencies that sum to one.  In a quantum pure state
$|\psi\rangle=\sum_s c_s|s\rangle$, the $c_s$ are complex amplitudes,
$\sum_s|c_s|^2=1$, and relative phases can change later measurement
probabilities.  Ordinary embeddings specify neither such phase-sensitive
transformations nor a Born-rule measurement.  Therefore a static convex
mixture is not a coherent quantum superposition, and contextualization is not
quantum collapse.

Appendix~\ref{app:polysemy} displays the pure-state and density-matrix
comparison explicitly.  Appendix~\ref{app:quantum-attention} instead supplies
a separate representational foil.  Its ``bank'' calculation shows how a
controlled relative phase can affect a Born-rule outcome after coherent
preparation and an incompatible readout.  The phase is not assigned a
linguistic meaning and is not part of ordinary transformer computation.

\subsection{A shared-projector incidence model}

A separate, limited analogy uses incidence rather than addition.  Here a
\emph{rank-one projective measurement context} means an orthonormal basis,
equivalently the associated mutually orthogonal rank-one projectors, which sum
to the identity.  In
dimension $d\geq3$, different contexts can share one projector while differing
in the others.

For the linguistic analogy, let each sense $s$ of ``bank'' have a context
$\mathcal E_s$ that contains the same lexical projector $P_{\rm bank}$.  The
other projectors specify the basis associated with $s$.  Here $s$ is a
linguistic sense and $\mathcal E_s$ is the quantum-style context assigned to
it, not the surrounding text.  This is a representation after $s$ has been
assigned, not a rule for inferring $s$.  Unlike
Eq.~\eqref{eq:static-mixture-main}, this model places one common element in
several bases; it does not average sense vectors.

\begin{figure}[t]
\centering
\begin{tikzpicture}[scale=0.5,
    outer/.style={circle, draw, minimum size=8pt,
                  inner sep=0pt, font=\footnotesize},
    lab/.style={font=\small}]
\coordinate (c) at (0,0);

\node[lab, above right=7pt] at (c)
{$P_{\rm bank}$};

\node[outer, red, fill=red!70,
      label={[lab]above:
      $P_{{\rm bank},{\rm finance},2}$}]
      (finance) at (90:4.2cm) {};

\node[outer, green, fill=green!70,
      label={[lab, xshift=-4pt]below left:
      $P_{{\rm bank},{\rm river},2}$}]
      (river) at (210:4.2cm) {};

\node[outer, blue, fill=blue!70,
      label={[lab, xshift=4pt]below right:
      $P_{{\rm bank},{\rm seat},2}$}]
      (seat) at (330:4.2cm) {};

\draw[very thick, red] (c) --
      node[lab, left=2pt, pos=0.45]
      {$\mathcal E_{\rm finance}$} (finance);

\draw[very thick, green!70!black] (c) --
      node[lab, above left=1pt, pos=0.45]
      {$\mathcal E_{\rm river}$} (river);

\draw[very thick, blue] (c) --
      node[lab, above right=1pt, pos=0.45]
      {$\mathcal E_{\rm seat}$} (seat);

\filldraw[blue, very thick] (c) circle (8pt);
\filldraw[green!70!black, very thick] (c) circle (6pt);
\filldraw[red, very thick] (c) circle (4pt);
\end{tikzpicture}
\caption{Shared-incidence schematic for ``bank.''  For each sense $s$, the
basis $\mathcal E_s$ contains the lexical projector $P_{\rm bank}$ and the
shown projector $P_{{\rm bank},s,2}$.  Unshown projectors complete each basis.
The diagram shows shared incidence, not a vector sum.}
\label{fig:contexts}
\end{figure}

Orthogonality has different roles in the two settings.  For an embedding
readout, $u\T v=0$ removes cross-talk from $v$ when the coefficient along $u$
is read by inner product; it does not imply statistical independence.  In
quantum theory, mutually orthogonal projectors can represent mutually
exclusive outcomes in one measurement context.  The analogy therefore uses
only the shared incidence structure.

The displayed configuration is Kochen--Specker colorable: assigning value
$1$ to $P_{\rm bank}$ and $0$ to every other projector gives exactly one value
$1$ in each context.  Shared incidence alone therefore does not establish
Kochen--Specker contextuality \cite{specker-60,kochen1}.

The shared projector does not select a sense; linguistic context may favor
one.  For a fixed quantum state, Appendix~\ref{app:polysemy} shows that the
Born probability of the shared projector does not depend on the other
projectors that complete its context.  Since any unit vector can be completed
to many bases, basis completion alone has no empirical content.  A direct test
would fix a layer, fit $r_t$ and $a_t$ on sense-labelled training occurrences,
and then freeze them.  On held-out occurrences, the RMS deviation of
$r_t\T h_\ell(t\mid x)$ from $a_t$ can be compared with sense-specific
coefficients and random directions.  Controlling for the initial lookup vector
would test whether the anchor is more than residual word-form identity.

\section{Relation to prior work}
\label{sec:prior-work}

Three adjacent literatures use related language for different objects.  In
mechanistic interpretability, \emph{feature superposition} means representing
more learned features than there are available dimensions; it does not imply
quantum coherence \cite{elhage2022superposition}.  Contextual-representation
and word-sense-disambiguation studies ask whether occurrence vectors
distinguish meanings in context
\cite{ethayarajh2019,pilehvar2019wic,wiedemann2019does,scarlini2020ares}; we
use that distinction but propose no new disambiguation method.  Quantum
cognition applies Hilbert-space probability to concepts and the mental lexicon
\cite{bruza2009lexicon,Busemeyer2012}; here it motivates a structural
comparison, not a claim that ordinary transformers implement quantum dynamics.

\section{Conclusion}
\label{sec:conclusion}

Searle's Chinese Room asks whether correct manipulation of symbols amounts to
understanding.  We have not answered that philosophical question.  We have
instead followed one operational path from tokens to meaning-sensitive
behavior.

Token identifiers first become vectors because vectors support graded
coordinates, learned transformations, comparison, and combination.  Their
high-dimensional geometry addresses the first problem considered here.  A
fixed-width representation space can accommodate exponentially many
quasi-orthogonal
directions at a chosen tolerance, even though it has only $d$ exactly
independent coordinates.  This is expressive room, not cost-free access.
Residual overlaps create readout cross-talk, especially when many relevant
features are active together.

Small overlaps reduce cross-talk in a linear semantic readout.  Environmental
overlaps can also suppress off-diagonal terms in a quantum reduced state, but
only the latter process is decoherence.  The two settings share an
inner-product pattern, not a physical mechanism.

The discussion of polysemy distinguishes four constructions: a static type
vector that may approximate a frequency-weighted convex mixture of sense
vectors; context-dependent occurrence vectors; a coherent quantum
superposition; and the optional shared-projector incidence construction.  The
first two are classical representation models, the third is a quantum
comparator rather than an ordinary transformer state, and the fourth does not
select a sense by itself.

At present, the relation between ordinary transformer LLMs and quantum mechanics
is best regarded as a limited analogy of inner-product geometry rather than a
common physical or semantic theory.  Language-model activations inhabit
finite-dimensional real inner-product spaces; standard quantum mechanics
represents pure states as rays in a complex Hilbert space and supplements this
geometry with observables and the Born rule; composite systems are represented
by tensor products.  Even shared terms have different
meanings: linguistic context is surrounding text, whereas a quantum
measurement context as used here is an orthonormal basis, equivalently its
rank-one projectors; ``interference''
here means linear readout cross-talk, whereas quantum interference is
phase-sensitive.

Appendix~\ref{app:quantum-attention} makes this distinction constructive.  Its
phase-sensitive result requires coherent complex amplitudes and a Born-rule
readout that ordinary transformer attention does not provide.  Because the
construction takes the scores as available and then compiles the required
state preparation, it is a representational foil rather than an end-to-end
quantum algorithm; no efficient implementation or speedup is established.

Quasi-orthogonality is nevertheless a genuine common geometric motif.  In a
language model it offers a possible capacity mechanism for many approximately
distinguishable features, subject to cross-talk and empirical confirmation.
In quantum theory, near-orthogonality of conditional environmental states can,
under appropriate dynamics, suppress reduced-state coherences and thereby
contribute to quasi-classical behavior.  It does not by itself select a pointer
basis, produce definite outcomes, or establish stable classical records.  The
paper thus identifies geometric resources for semantic discrimination, not a
complete theory of machine understanding.

\appendix

\section{Formal companion to \emph{From text to vectors}}
\label{app:vectors}

This appendix gives a minimal mathematical version of
Sec.~\ref{sec:vectors}.  It distinguishes strings, token identifiers, and
vectors before discussing semantic directions.

\subsection{Tokenization and lookup}

Let $\mathcal S$ be the set of finite character strings.  Let
$\mathcal V=\{1,\ldots,V\}$ be a vocabulary containing $V$ token identifiers.
Each identifier is an integer label.  A tokenizer is a fixed map
\begin{equation}
 \mathsf{Tok}:\mathcal S\longrightarrow
 \bigcup_{T\geq0}\mathcal V^T.
 \label{eq:tokenizer-app}
\end{equation}
If $\mathsf{Tok}(\sigma)=(t_1,\ldots,t_T)$, then the input string $\sigma$
has become a sequence of $T$ identifiers.  The union in
Eq.~\eqref{eq:tokenizer-app} allows strings to produce sequences of different
lengths.  A detokenizer reverses the assembly, subject to the system's stated
normalization rules.

Let $d_{\rm model}$ be the architectural width of the model's hidden stream.
The notation $\R^m$ means the set of real column vectors with $m$ coordinates.
An embedding table is a real matrix
$E\in\R^{V\times d_{\rm model}}$.  Its row $E[t_i]$ is the initial lookup
vector for identifier $t_i$.  In a simple additive positional scheme,
\begin{equation}
 h_i^{(0)}=E[t_i]+p_i,
 \qquad h_i^{(0)},p_i\in\R^{d_{\rm model}}.
 \label{eq:lookup-app}
\end{equation}
Here $p_i$ supplies the position and the superscript $(0)$ means the stage
before the first transformer layer.  Some architectures incorporate position
inside later comparison operations rather than literally adding $p_i$; the
conceptual role is the same.  The identifier $t_i$, the table row $E[t_i]$, and the
contextual state produced at a later layer are three different objects.

The geometric analysis need not use all $d_{\rm model}$ architectural
coordinates.  A declared preprocessing map may center the observed states and
remove nearly unused directions.  It may also rescale the retained principal
directions to a declared common variance, a step called whitening.  We use
$d\leq d_{\rm model}$ for the rank of the resulting representation space.
Any numerical claim involving $d$ must therefore report the model,
layer, sample, centering, rank threshold, and inner product used to obtain it.

\subsection{Projection onto a fitted feature direction}

Let $u\in\R^d$ be a unit vector, so $u\T u=1$.  The orthogonal projection of a
state $h$ onto the line through $u$ is
\begin{equation}
 P_u h=(u\T h)u,
 \qquad P_u=uu\T .
 \label{eq:projection-app}
\end{equation}
The scalar $u\T h$ is its signed coordinate.  The matrix $P_u$ is called a
projector because direct multiplication gives $P_u^2=P_u$.  The remainder
$h-P_u h$ is perpendicular to $u$, since
$u\T(h-P_u h)=0$.  This is the elementary orthogonal projection theorem in
finite-dimensional form.

An oriented contrast needs a baseline.  Suppose two training samples have
means $\mu_A$ and $\mu_B$ in the same representation space and
$\mu_A\ne\mu_B$.  Define
\begin{equation}
\begin{aligned}
 g&=\frac{\mu_A-\mu_B}{\norm{\mu_A-\mu_B}},
 &h_0&=\frac{\mu_A+\mu_B}{2},\\
 \alpha_g(h)&=g\T(h-h_0).
\end{aligned}
 \label{eq:signed-axis-app}
\end{equation}
The unit vector $g$ orients the contrast, $h_0$ declares its midpoint, and
$\alpha_g(h)$ is the signed coordinate.  Substitution shows
\begin{equation}
 \alpha_g(\mu_A)=\frac{\norm{\mu_A-\mu_B}}{2},
 \qquad
 \alpha_g(\mu_B)=-\frac{\norm{\mu_A-\mu_B}}{2}.
 \label{eq:mean-scores-app}
\end{equation}
Swapping the two groups reverses $g$ and every sign but does not change the
unoriented line.  A score near zero means that the coordinate along $g$ is near
the midpoint; it does not imply that the full vector is near $h_0$, because a
large perpendicular component may remain.

The same construction can illustrate a fitted gender-associated contrast,
but it does not define gender and need not capture all gender-related
information.  A linear classifier provides a related alternative: if its
decision boundary is $w\T h+\beta_0=0$ with $w\ne0$, then
$(w\T h+\beta_0)/\norm w$ is signed Euclidean distance from that hyperplane.
Either construction must be fitted on training examples and evaluated on
held-out examples.

\section{Formal companion to the geometric argument}
\label{app:geometry}

The main text used three counts.  The integer $d$ is the exact dimension of
the selected vector space.  The integer $M$ is the number of feature or code
directions in a declared dictionary.  The integer $k$ is the number of those
directions appreciably active in one state.  Only $d$ is a vector-space
dimension.

\subsection{Exact orthogonality and quasi-orthogonality}

Let $w_1,\ldots,w_M\in\R^d$ be unit vectors.  The family is
$\varepsilon$-quasi-orthogonal when
\begin{equation}
 |w_i\T w_j|\leq\varepsilon
 \quad\hbox{whenever }i\ne j,
 \qquad 0<\varepsilon<1.
 \label{eq:quasi-app}
\end{equation}
The absolute value treats a direction and its negative as the same
unoriented axis for purposes of overlap.

When $\varepsilon=0$, the vectors are mutually orthogonal.  To see why there
can then be at most $d$, suppose
$a_1w_1+\cdots+a_Mw_M=0$.  Taking the inner product with $w_j$ leaves
$a_j=0$, because every cross-term vanishes and $w_j\T w_j=1$.  All
coefficients are zero, so the vectors are linearly independent.  The
definition of dimension now gives $M\leq d$.

For $\varepsilon>0$, the vectors need not be linearly independent.  The
largest possible $M$ becomes a finite spherical-code problem at the chosen
tolerance.  Its cardinality is sometimes informally called
quasi-dimensionality, but no additional linear dimensions have appeared.

\subsection{Why typical overlaps shrink with dimension}

For a random quantity $Y$, $\E[Y]$ denotes its average.  For an event $A$,
$\Prob(A)$ denotes its probability.  A unit vector chosen uniformly from the
sphere has no preferred direction: rotating the whole experiment does not
change its probability law.

Fix a unit vector $u$ and choose $X$ uniformly from the unit sphere in
$\R^d$.  Rotational symmetry lets us take $u$ to be the first coordinate
direction.  The overlap $u\T X$ is then $X_1$.  Symmetry makes all coordinate
squares have the same expectation.  Since
$X_1^2+\cdots+X_d^2=1$, taking expectations gives
\begin{equation}
 1=d\,\E[X_1^2],
 \qquad
 \E[(u\T X)^2]=\frac1d.
 \label{eq:rms-overlap-app}
\end{equation}
The RMS overlap is the square root of the mean squared overlap, hence
$1/\sqrt d$.  This elementary calculation gives a scale, not a tail
probability and not a guarantee for learned vectors.

L\'evy's concentration lemma supplies the tail.  One convenient consequence
is that universal positive constants $c$ and $C$ exist such that
\begin{equation}
 \Prob\{|u\T X|>\varepsilon\}
 \leq C\exp(-c d\varepsilon^2)
 \label{eq:levy-overlap-app}
\end{equation}
for a fixed unit $u$ and a relevant fixed tolerance.  The function
$x\mapsto u\T x$ is 1-Lipschitz because the Cauchy--Schwarz inequality gives
$|u\T(x-y)|\leq\norm{x-y}$.  L\'evy's lemma says that a Lipschitz function on
a high-dimensional sphere is unlikely to deviate far from its typical value.
Here 1-Lipschitz means that changing the input by a distance $r$ changes the
function value by at most $r$
\cite{ledoux2001concentration,milman1986asymptotic}.

\subsection{From one pair to an exponential family}

Choose $M$ directions independently and uniformly.  There are
$M(M-1)/2$ unordered pairs.  The union bound states that the probability of at
least one bad pair is no larger than the sum of the bad-pair probabilities.
Combining it with Eq.~\eqref{eq:levy-overlap-app} gives
\begin{equation}
 \Prob\!\left\{
 \max_{i\ne j}|w_i\T w_j|>\varepsilon
 \right\}
 \leq \frac{C M(M-1)}{2}
 \exp(-c d\varepsilon^2).
 \label{eq:union-app}
\end{equation}
Choose the greatest integer $M$ not exceeding
$\exp(c'\varepsilon^2d)$, where $c'$ is sufficiently small and $d$ is
sufficiently large.  Then the right side is below one.  Therefore at least one
family of that exponential size has every overlap within tolerance.  This is
an existence lower bound and a random-code benchmark, not a universal maximum
and not proof that an LLM learns such a code.

Solving the exponent in Eq.~\eqref{eq:union-app} for the overlap tolerance
shows that the union-bound crossover is a constant multiple of
$\sqrt{\log M/d}$ when $\log M$ is small compared with $d$.  The union bound
supplies a high-probability upper scale, not by itself a matching lower law.
This remains a null-model estimate rather than a hard capacity wall.

\subsection{The Johnson--Lindenstrauss route}

The Johnson--Lindenstrauss lemma states that a set of $M+1$ points can be mapped
into $\R^d$ while preserving all squared distances within relative error
$\delta$ when
\begin{equation}
 d\geq C_{\rm JL}\delta^{-2}\log(M+1),
 \qquad 0<\delta<1,
 \label{eq:jl-app}
\end{equation}
where $C_{\rm JL}$ is a numerical constant and $\log$ is the natural logarithm
\cite{johnson-1984,Dasgupta-2003}.  This is a theorem about a prescribed finite
set, not random semantic features.

To obtain a quasi-orthogonal code, take the origin and $M$ mutually orthogonal
standard coordinate vectors $e_i$ in a sufficiently large starting space.
Let $\Phi$ be a linear Johnson--Lindenstrauss map and define
$y_i=\Phi e_i$.  Including the origin controls the image lengths.  The
polarization identity
\begin{equation}
 2y_i\T y_j=\norm{y_i}^2+\norm{y_j}^2-\norm{y_i-y_j}^2
 \label{eq:polarization-app}
\end{equation}
turns preservation of lengths and distances into preservation of inner
products.  Before normalization their pairwise absolute inner products are at
most $2\delta$; because every squared length is at least $1-\delta$, the
normalized overlaps are at most $2\delta/(1-\delta)$.  For a fixed target
dimension $d$, choose an integer $M$ satisfying
$M+1\leq\exp(c\delta^2d)$, with $c$ small enough that
Eq.~\eqref{eq:jl-app} holds.  This constructs an exponential-size family.

L\'evy concentration, random spherical coding, and the
Johnson--Lindenstrauss lemma are thus related views of one high-dimensional
phenomenon.  Their exponential factors must not be multiplied as though they
were independent mechanisms.

\subsection{A deterministic overlap floor}

Random coding gives a typical construction.  The Welch bound gives a
deterministic limitation.  Put the vectors into the columns of a matrix
$W\in\R^{d\times M}$ and form the Gram matrix $G=W\T W$.  Its entry
$G_{ij}=w_i\T w_j$ records every pairwise overlap.  The diagonal entries are
one, so the trace of $G$ is $M$.  The matrix has rank at most $d$ and
nonnegative eigenvalues.

Let $\lambda_1,\ldots,\lambda_r$ be its nonzero eigenvalues, where $r\leq d$.
Cauchy--Schwarz gives
\begin{equation}
 M^2=\left(\sum_{\ell=1}^r\lambda_\ell\right)^2
 \leq r\sum_{\ell=1}^r\lambda_\ell^2
 \leq d\sum_{\ell=1}^r\lambda_\ell^2.
 \label{eq:eigen-cs-app}
\end{equation}
The sum of squared matrix entries equals the sum of squared eigenvalues for a
real symmetric matrix.  Removing the $M$ diagonal ones and averaging the
remaining entries yields
\begin{equation}
 \max_{i\ne j}|w_i\T w_j|^2
 \geq \frac{M-d}{d(M-1)}
 \qquad (M>d).
 \label{eq:welch-app}
\end{equation}
This is the Welch bound \cite{welch1974coherence}.  It says that once the
dictionary is overcomplete, some nonzero overlap is unavoidable.  It does not
say which pair overlaps or how damaging that pair is for a particular task.

\subsection{Two-feature and many-feature readout}

Let a hidden state be represented by
\begin{equation}
 h=\sum_{i\in A}a_iw_i+e.
 \label{eq:feature-code-app}
\end{equation}
Here $A$ is the active set, $a_i$ are scalar feature coefficients, and $e$ contains
components omitted by this chosen dictionary.  Reading feature $j\in A$ by
correlation gives
\begin{equation}
 w_j\T h
 =a_j+\sum_{\substack{i\in A\\i\ne j}}
 a_i(w_j\T w_i)+w_j\T e.
 \label{eq:readout-app}
\end{equation}
The first term is desired; the sum contains geometric cross-talk.
Equation~\eqref{eq:two-feature-main} is the case $A=\{1,2\}$ and $e=0$.

If the directions are $\varepsilon$-quasi-orthogonal, the triangle inequality
gives the worst-case bound
\begin{equation}
 \left|
 \sum_{\substack{i\in A\\i\ne j}}a_i(w_j\T w_i)
 \right|
 \leq\varepsilon\sum_{\substack{i\in A\\i\ne j}}|a_i|.
 \label{eq:worst-readout-app}
\end{equation}
This bound allows every unwanted term to have the same sign.  Under a
different, explicitly random-like model, assume instead that the cross-terms
are centered, have zero cross-moments, and satisfy
$\E[(w_j\T w_i)^2\mid w_j]=1/d$.  Treat the active set and coefficients as fixed
and condition on $w_j$.  Variances then add:
\begin{equation}
 \E\!\left[
 \left.
 \left(\sum_{\substack{i\in A\\i\ne j}}a_iw_j\T w_i\right)^2
 \right|A,(a_i),w_j
 \right]
 =\frac1d\sum_{\substack{i\in A\\i\ne j}}a_i^2.
 \label{eq:variance-readout-app}
\end{equation}
For $k-1$ interfering features of comparable unit coefficient, the square root
is $\sqrt{(k-1)/d}$.  This is the RMS law quoted in the main text.  Centering
or zero cross-moments is essential; an informal claim of uncorrelated
contributions would be insufficient without their means being defined.

The contrast between Eqs.~\eqref{eq:worst-readout-app} and
\eqref{eq:variance-readout-app} explains why no universal Goldilocks boundary
follows from $d$ and $M$ alone.  A task must also specify active coefficients,
coactivation frequencies, a decoder, and an accepted error.  Overlap between
two features that never need to be recovered together may have little cost.

\subsection{Linear readout cross-talk and quantum decoherence}
\label{app:dirac}

We first introduce the notation used in this comparison.  A pure quantum
state may be represented by a normalized complex column vector; vectors that
differ only by an overall phase represent the same physical ray.  Dirac
notation writes such a vector as a
\emph{ket} $|v\rangle$ and its conjugate transpose as the \emph{bra}
$\langle v|$.  Thus $\langle v|w\rangle$ is an inner product, whereas
$|v\rangle\langle w|$ is a matrix called an outer product.  A density matrix
is a positive semidefinite matrix of trace one; the trace is the sum of its
diagonal entries.  The partial trace over an environment is the linear
operation that discards the environmental coordinates while preserving all
expectation values of measurements made only on the system.

For comparison with physical decoherence, let
$\{|s_i\rangle\}$ be orthonormal alternatives of a quantum system.  Suppose a
system--environment interaction produces the pure joint state
\begin{equation}
 |\Psi\rangle=\sum_i c_i\,|s_i\rangle|E_i\rangle,
 \qquad \sum_i|c_i|^2=1.
 \label{eq:branch-state-app}
\end{equation}
The normalized $|E_i\rangle$ are conditional environmental states.  Observing
only the system means taking a partial trace over the environment.  The
reduced system state is
\begin{equation}
 \rho_S
 =\sum_{i,j}c_i c_j^*
 \langle E_j|E_i\rangle\,|s_i\rangle\langle s_j|.
 \label{eq:reduced-state-app}
\end{equation}
Thus every off-diagonal term with $i\ne j$ is multiplied by an environmental
overlap.  When the relevant $|E_i\rangle$ become nearly orthogonal, those
coherence terms become small for system-only observations.  This is the
standard algebraic core of environmental decoherence
\cite{RevModPhys.75.715}.

Near-orthogonality of the $|E_i\rangle$ requires a stated assumption about the
interaction or branch-state ensemble; high dimension alone does not produce
it.  Moreover, many small off-diagonal terms can accumulate.  Therefore,
bounding each pair separately does not by itself show that the whole reduced
matrix is close to the matrix obtained by deleting all off-diagonal terms.

Compare Eq.~\eqref{eq:reduced-state-app} with
Eq.~\eqref{eq:readout-app}.  The quantum coherence term is suppressed by
$\langle E_j|E_i\rangle$; the classical readout error is suppressed by
$w_j\T w_i$.  Both are small-overlap effects, but only the first follows from
a joint quantum state and a partial trace.  In the LLM setting the effect is
ordinary suppression of linear-readout cross-talk, not decoherence.

\subsection{Conditional tensor-product amplification}

One further scaling is mathematically correct but not implied for an ordinary
transformer.  If a separate model genuinely has $n$ tensor factors, each of
fixed local dimension $q\geq2$, then its exact Hilbert-space dimension is
\begin{equation}
 d=q^n.
 \label{eq:tensor-dimension-app}
\end{equation}
Substitution into the exponential code bound gives
\begin{equation}
 M\geq\exp(c\varepsilon^2q^n)-1,
 \label{eq:double-exponential-app}
\end{equation}
which is doubly exponential in the number of factors $n$ at fixed $q$ and
$\varepsilon$, for sufficiently large $n$.  This is a code cardinality, not
an additional Hilbert-space dimension.  A real code is also a valid subset of
the corresponding complex representation space.  Standard LLM residual width
is not known to equal an accessible $q^n$ tensor-product representation space, so
Eq.~\eqref{eq:double-exponential-app} is not an ordinary LLM capacity claim.

\section{Formal companion to the polysemy comparison}
\label{app:polysemy}

\subsection{Static and contextual vectors}

Let $t$ be a token type that represents a complete word, and let
$s\in\{1,\ldots,S_t\}$ index that word's senses.  Under the generative model
of Arora et al., the population type vector $\bar v_t$ is approximately
\begin{equation}
\begin{aligned}
 \bar v_t&\approx\sum_{s=1}^{S_t}\pi_t(s)v_{t,s},\\
 \pi_t(s)&=\frac{f_{t,s}}{\sum_r f_{t,r}},
 &\qquad \sum_s\pi_t(s)&=1.
\end{aligned}
 \label{eq:arora-app}
\end{equation}
Here $v_{t,s}$ is a hypothetical sense vector and $f_{t,s}$ is the number of
occurrences with sense $s$.  This relation is model-dependent.  Arora et al.\
prove the two-sense case explicitly; the displayed multisense form uses the
same conditioning argument
\cite{arora-etal-2018-linear}.

A transformer makes the different statement that a learned map $F_\ell$
produces one state for an occurrence in linguistic context $x$:
\begin{equation}
 h_\ell(t\mid x)=F_\ell(E[t],x)
 \label{eq:context-state-app}
\end{equation}
at layer $\ell$, where $E[t]$ is the token-type lookup vector.  This map need
not be a convex mixture of fixed sense vectors.

\subsection{Quantum pure superposition and incoherent mixture}

Using the Dirac notation introduced in Appendix~\ref{app:dirac}, let
$\{|s\rangle\}$ now denote orthonormal quantum alternatives.  A normalized
pure state is
\begin{equation}
 |\psi\rangle=\sum_s c_s|s\rangle,
 \qquad
 \sum_s|c_s|^2=1.
 \label{eq:pure-superposition-app}
\end{equation}
The coefficients $c_s$ are amplitudes.  Their relative signs or phases can
affect probabilities after a change of measurement basis.

The density matrix of the pure state is
\begin{equation}
 |\psi\rangle\langle\psi|
 =\sum_{s,r}c_s c_r^*|s\rangle\langle r|.
 \label{eq:pure-density-app}
\end{equation}
The terms with $s\ne r$ are phase-sensitive coherences.  By contrast, an
incoherent mixture of the same quantum alternatives is
\begin{equation}
 \rho_{\rm mix}=\sum_s p_s|s\rangle\langle s|,
 \qquad p_s\geq0,\quad\sum_s p_s=1.
 \label{eq:density-mixture-app}
\end{equation}
Equation~\eqref{eq:density-mixture-app} has no off-diagonal terms in this
basis.  The lexical average in Eq.~\eqref{eq:arora-app} is neither
Eq.~\eqref{eq:pure-superposition-app} nor
Eq.~\eqref{eq:density-mixture-app}: it is an average of real embedding
vectors, with no quantum measurement rule.

\subsection{One projector shared by several sense-labelled contexts}

Let $d\geq3$, let $r_t\in\R^d$ be a unit lexical anchor, and put
$P_t=r_t r_t\T$.  For each sense $s$, choose an orthonormal basis
\begin{equation}
 \{r_t,u_{t,s,2},\ldots,u_{t,s,d}\}
 \label{eq:residual-basis-app}
\end{equation}
and define $P_{t,s,j}=u_{t,s,j}u_{t,s,j}\T$.  The associated quantum-style
context is
\begin{equation}
 \mathcal E_{t,s}
 =\{P_t,P_{t,s,2},\ldots,P_{t,s,d}\},
 \qquad
 P_t+\sum_{j=2}^{d}P_{t,s,j}=I_d.
 \label{eq:context-hyperedge-app}
\end{equation}
For the quantum-style reading, these real vectors and projectors are identified
with their canonical embeddings in $\mathbb C^d$.
Different sense-labelled contexts can share $P_t$ and differ in their other
projectors.  Completing $r_t$ to such bases is automatic; it is not an
empirical result.  Because each basis spans the same space, its choice alone
adds no constraint; the fixed projection below is the substantive part.
The resulting star of bases is Kochen--Specker colorable: setting
$v(P_t)=1$ and assigning value $0$ to every other projector gives exactly one
value $1$ in each context.  Thus this incidence pattern alone is not a proof
of quantum contextuality.

An idealized shared-anchor hypothesis for occurrence states can be written
\begin{equation}
 h_\ell(t\mid x)
 =a_t r_t+\sum_{j=2}^{d}z_{t,s(x),j}(x)u_{t,s(x),j} .
 \label{eq:shared-anchor-app}
\end{equation}
Here $x$ is the linguistic context and $s(x)$ is an assigned sense label; the
equation does not infer that label.  Its exact form requires the same
nonzero $a_t$ for every occurrence of type $t$.  This common coefficient is
the testable hypothesis; completing $r_t$ to the bases is automatic.

A direct test fixes a layer $\ell$, estimates $r_t$ and $a_t$ on
sense-labelled training occurrences, and then freezes them.  On held-out
occurrences, report the RMS error of $r_t\T h_\ell(t\mid x)-a_t$ and compare it
with a model that allows a separate coefficient $a_{t,s}$ for each sense and
with matched random directions.  The shared-anchor hypothesis requires a
nonzero common coefficient with little held-out penalty relative to the
sense-specific model.  Repeating the analysis after controlling for the
initial lookup vector separates a learned anchor from persistent type identity.

For a fixed quantum state $\rho$, the Born probability assigned to the shared
projector is
\begin{equation}
 p(P_t\mid\rho)=\operatorname{tr}(\rho P_t).
 \label{eq:born-shared-app}
\end{equation}
No term in this expression names the other projectors that complete the
measurement context.  Hence the shared projector alone does not select a
sense.


\section{A minimal quantum attention model as a representational foil}
\label{app:quantum-attention}

This appendix gives a minimal quantum attention construction as a
\emph{representational foil}.  Its purpose is to display the extra mathematical
structure used here to make relative phase observable: a complex Hilbert
space, coherent unitary state preparation, a tensor-product address--value
state, and projective Born readout.  The partial-trace output supplies the
incoherent comparison.  The construction is neither a reinterpretation of
ordinary transformer attention nor a proposed efficient quantum algorithm.
No quantum speedup or empirical linguistic advantage is claimed.

For each target $i$, the construction uses an address space
$\mathcal H_A^{(i)}\simeq\mathbb C^i$ with orthonormal basis
$\{|j\rangle_A:j=1,\ldots,i\}$ and a semantic or value space
$\mathcal H_S\simeq\mathbb C^V$ with a complete orthonormal vocabulary basis
$\{|y\rangle_S:y=1,\ldots,V\}$, where $V$ is the vocabulary size.  The
complete orthonormal vocabulary basis is
an explicit assumption of this toy model.  Their joint space is
$\mathcal H_A^{(i)}\otimes\mathcal H_S$; tracing out the address factor leaves
a state on $\mathcal H_S$.

For each $i$ choose a normalized reference $|0\rangle_A\in\mathcal H_A^{(i)}$
and choose $|0\rangle_S\in\mathcal H_S$.  Encode source position $j$ and form
query, key, and value states by unitary matrices:
\begin{equation}
\begin{aligned}
 |\psi_j\rangle&=U_{\rm enc}(x_j)|0\rangle_S,\\
 |q_i\rangle&=Q_\theta|\psi_i\rangle,\qquad
 |k_j\rangle=K_\theta|\psi_j\rangle,\qquad
 |v_j\rangle=V_\theta|\psi_j\rangle .
\end{aligned}
\label{eq:mqa-encoding}
\end{equation}
Here $Q_\theta,K_\theta,V_\theta\in U(V)$ are learned.  The complex
query--key overlap is the score.  Independently formed overlaps are not
generally one row of a unitary matrix, so they must be normalized explicitly:
\begin{equation}
\begin{aligned}
 c_{ij}&=\langle q_i|k_j\rangle,
 &Z_i&=\sum_{r\leq i}|c_{ir}|^2,\\
 \alpha_{ij}&=\frac{c_{ij}}{\sqrt{Z_i}},
 &Z_i&>0 .
\end{aligned}
\label{eq:mqa-scores}
\end{equation}
Thus $\sum_{j\leq i}|\alpha_{ij}|^2=1$.  The $\alpha_{ij}$ are amplitudes
rather than classical weights, and their relative phases are retained.

Let one prefix-dependent unitary $W_i=W_i(x_{\leq i};\theta)$ prepare
\begin{equation}
 |\Omega_i\rangle
 =W_i|0\rangle_A|0\rangle_S
 =\sum_{j\leq i}\alpha_{ij}|j\rangle_A|v_j\rangle_S .
\label{eq:mqa-unitary}
\end{equation}
The right-hand side is normalized because the address states are orthogonal
and every value state is normalized.  It can therefore be extended to an
orthonormal basis and chosen as one column of a unitary matrix.
Operationally, $W_i$ may be compiled as address-state preparation followed by
token-controlled value preparation.  In this specification, however, the
scores $c_{ij}$ and normalized amplitudes $\alpha_{ij}$ must be obtained before
$W_i$ is compiled.  No coherent score-computation routine or efficient
state-preparation circuit is given.

A projective measurement of the address gives the normalized Born attention
weights, without a softmax:
\begin{equation}
 p_i(j)
 =\operatorname{tr}\!\left[
 (|j\rangle\langle j|_A\otimes I_S)
 |\Omega_i\rangle\langle\Omega_i|
 \right]
 =|\alpha_{ij}|^2
 =\frac{|c_{ij}|^2}{Z_i}.
\label{eq:mqa-address}
\end{equation}
If the address is traced out---whether or not it is first measured and
forgotten or completely dephased---the retained value state is simply
\begin{equation}
 \rho_{S,i}^{\rm mix}
 =\operatorname{tr}_A|\Omega_i\rangle\langle\Omega_i|
 =\sum_{j\leq i}p_i(j)|v_j\rangle\langle v_j|.
\label{eq:mqa-mixture}
\end{equation}
Equation~\eqref{eq:mqa-mixture} defines the incoherent output of the model.
It contains no interference between different addresses.  In
particular, the partial trace does not produce the coherent sum
$\sum_j\alpha_{ij}|v_j\rangle$.

From $|\Omega_i\rangle$, that sum is obtained by a postselected
complementary-basis projective readout rather than by tracing out $A$.  Put
\begin{equation}
 |+\rangle_A=\frac{1}{\sqrt{i}}\sum_{j\leq i}|j\rangle_A,\qquad
 N_i=\left\|\sum_{j\leq i}\alpha_{ij}|v_j\rangle\right\|^2 .
\label{eq:mqa-erasure}
\end{equation}
When $N_i>0$, conditioned on the $|+\rangle_A$ outcome, the semantic state and the
probability of obtaining it are
\begin{equation}
 |\phi_i\rangle
 =\frac{\sum_{j\leq i}\alpha_{ij}|v_j\rangle}{\sqrt{N_i}},
 \qquad
 r_i=\Pr(+)=\frac{N_i}{i}.
\label{eq:mqa-coherent}
\end{equation}
This postselected projective ``quantum eraser'' removes which-address information.
Measuring which address was selected and coherently erasing that information
are alternative readouts of the same prepared state.

The orthonormal vocabulary basis now gives an ordinary projective Born rule:
\begin{equation}
\begin{aligned}
 p_i^{\rm coh}(y\mid +)
 &=|\langle y|\phi_i\rangle|^2
 =\frac{\left|\sum_{j\leq i}
       \alpha_{ij}\langle y|v_j\rangle\right|^2}{N_i},\\
 p_i^{\rm mix}(y)
 &=\operatorname{tr}\!\left(
 |y\rangle\langle y|\rho_{S,i}^{\rm mix}\right)
 =\sum_{j\leq i}p_i(j)|\langle y|v_j\rangle|^2 .
\end{aligned}
\label{eq:mqa-vocabulary}
\end{equation}
Both distributions sum to one.  The coherent probability contains
phase-sensitive cross-terms; the mixture does not.  The parameters of the
unitary matrices may be trained by minimizing, for example,
$-\sum_i\log p_i^{\rm coh}(y_i^{\rm true}\mid +)$.  A physical implementation
must also account for the erasure success probability $r_i$, for instance by
training the joint probability $r_i p_i^{\rm coh}(y_i^{\rm true}\mid +)$.

For a diagnostic phase intervention, take two orthonormal value states
$|{\rm F}\rangle$ and $|{\rm R}\rangle$, labelled by the financial and
river-edge senses of ``bank.''  Let
$\alpha_{\rm F}=1/\sqrt2$ and
$\alpha_{\rm R}=e^{i\varphi}/\sqrt2$.  Here $\varphi$ is only a control
parameter: the construction gives it no linguistic interpretation and no rule
by which text would set or learn it.  The orthogonality of the two
sense-labelled states is likewise an assumption of this toy example.
Successful address erasure prepares
\begin{equation}
 |\phi_\varphi\rangle
 =\frac{|{\rm F}\rangle+e^{i\varphi}|{\rm R}\rangle}{\sqrt2}.
\label{eq:mqa-bank-state}
\end{equation}
Within this two-dimensional subspace, use the readout basis
$|\mathord{\pm}\rangle=(|{\rm F}\rangle\pm|{\rm R}\rangle)/\sqrt2$.
The projective Born probabilities are
\begin{equation}
 p_+(\varphi)=\cos^2\frac{\varphi}{2},\qquad
 p_-(\varphi)=\sin^2\frac{\varphi}{2}.
\label{eq:mqa-bank-fringe}
\end{equation}
Changing $\varphi$ from zero to $\pi$ exchanges the two readout probabilities,
while the direct $|{\rm F}\rangle,|{\rm R}\rangle$ populations remain $1/2$.
By contrast, Eq.~\eqref{eq:mqa-mixture} gives the phase-insensitive state
$(|{\rm F}\rangle\langle{\rm F}|
 +|{\rm R}\rangle\langle{\rm R}|)/2$ and therefore $p_+=p_-=1/2$.
Thus relative phase changes a prediction while direct-path populations are
held fixed.

These equations can be simulated classically.  They specify states and
readouts, not an end-to-end quantum attention algorithm.  In particular, no
coherent score-computation routine or efficient state-preparation procedure is
provided, so the appendix establishes no quantum computational advantage.  If
instantiated on quantum hardware, $W_i$, address erasure, and projective
measurement would be physical operations subject to ordinary measurement and
no-cloning constraints.  The required states would have to be freshly
prepared from classically specified tokens; no unknown quantum state is
copied.

\begin{acknowledgments}
OpenAI Codex using GPT-5 (generative pre-trained transformer, version 5) was
used to assist with literature organization, \LaTeX{}
editing, and checks of derivations.  The author specified the scientific
direction and prompts; model output retained in the manuscript was checked
against the cited sources and explicit calculations.  The author accepts full
responsibility for the final text.

This research was funded in part by the Austrian Science Fund (FWF), grant
digital object identifier (DOI)
\href{https://doi.org/10.55776/PIN5424624}{10.55776/PIN5424624}.
The author also acknowledges support from the TU Wien Bibliothek Open Access
Funding Programme.

The author declares no conflict of interest.
\end{acknowledgments}

\bibliography{svozil}

\end{document}